\documentclass{article}
\usepackage{spconf,amsmath,graphicx,amssymb,setspace}

\title{Distributions of Upper PAPR and Lower PAPR of OFDM Signals\\ in Visible Light Communications}
\name{Zhenhua Yu$^{\star}$ \qquad Robert J. Baxley$^{\dag}$ \qquad G. Tong Zhou$^{\star}$ \thanks{This research was supported in part by the Texas Instruments Leadership University Program.}}

\address{$^{\star}$ School
of Electrical and Computer Engineering, Georgia Institute of Technology, Atlanta,
GA, 30332 USA \\
    $^{\dag}$ Georgia Tech Research Institute, Atlanta,
    GA, 30332 USA}
\begin{document}
%\ninept
%
\maketitle
\begin{abstract}
Orthogonal frequency-division multiplexing (OFDM) in visible light communications (VLC) inherits the disadvantage of high peak-to-average power ratio (PAPR) from OFDM in radio frequency (RF) communications. The upper peak power and lower peak power of real-valued VLC-OFDM signals are both limited by the dynamic constraints of light emitting diodes (LEDs). 
The efficiency and transmitted electrical power are directly related with the upper PAPR (UPAPR) and lower PAPR (LPAPR) of VLC-OFDM. In this paper, we will derive the complementary cumulative distribution
function (CCDF) of UPAPR and LPAPR, and investigate the joint distribution of UPAPR and LPAPR.
\end{abstract}
\section{Introduction}
\label{sec:intro}
 Motivated by the rapid progress of solid state lighting technology and increasingly saturated radio frequency (RF) spectrum, visible light communication (VLC) has become a promising candidate to complement conventional RF communication \cite{O'Brien2008,Elgala2011,Noshad2013}. VLC uses the visible light spectrum to provide illumination and communication simultaneously by way of light emitting diodes (LEDs) \cite{vlcstandard}. In VLC, simple and low-cost intensity modulation and direct detection (IM/DD) techniques are employed, thus only signal intensity information, not phase information, is modulated.

 Recently, orthogonal frequency-division multiplexing (OFDM) has been considered for VLC due to its ability to boost data rates and effectively combat inter-symbol-interference (ISI) \cite{Elgala2007, Armstrong2009, Yu2012a}. OFDM is also known for its disadvantage of high peak-to-average power ratio (PAPR). Power amplifiers in RF communication systems often have to operate with large power back-off and reduces the power efficiency \cite{baxley2009}. The distribution of PAPR of complex-valued RF-OFDM baseband signals has been extensively studied in references \cite{Bauml1996a,Ochiai2001,Goeckel2002b,Jiang2008}.  VLC-OFDM inherits the high PAPR from RF-OFDM. However, different from RF-OFDM, VLC-OFDM baseband signals must be real-valued required by IM/DD schemes. Thus, the frequency-domain symbols of OFDM must be Hermitian symmetric to make the time-domain samples real-valued. Additional, rather than peak power constrained in RF-OFDM, VLC-OFDM is dynamic range constrained by the turn-on current and maximum permissible alternating current of LEDs \cite{Elgala2009}. Furthermore, illumination requirements place a constraint on the average amplitude of the VLC-OFDM signal. For the real-valued VLC-OFDM signal, the square of the maximum value can be seen as the upper peak power, and the square of the minimum value can be seen as the lower peak power. Upper peak and lower peak have individual constraints in VLC. It has been studied in references \cite{Yu2013,Yu2013a} that the efficiency and transmitted electrical power are directly related with the upper PAPR (UPAPR) and lower PAPR (LPAPR) of VLC-OFDM. Although the distribution of PAPR of real-valued OFDM was shown in reference \cite{Yu2003,Ma2012d}, to the best of our knowledge, the distribution of UPAPR and LPAPR are still unknown.

In this paper, we will derive the complementary cumulative distribution
function (CCDF) of UPAPR and LPAPR, and the joint distribution of UPAPR and LPAPR, assuming that VLC-OFDM time-domain signals are independent and identically Gaussian distributed for large number of subcarriers. Simulated results are provided to examine our theoretical analysis. 

\section{Review of PAPR in RF-OFDM }
In RF-OFDM systems, let $\{X_k\}_{k=0}^{N-1}$ denote the frequency-domain OFDM signal, where $k$ is the subcarrier index and $N$ is the number of subcarriers. The Nyquist-rate sampled time-domain OFDM signal $\{x[n]\}_{n=0}^{N-1}$ is generated by applying inverse discrete Fourier transform (IDFT) to the frequency-domain signal:
\begin{equation}
\label{eq:OFDM}
x[n] = \frac{1}{\sqrt{N}}\sum_{k=0}^{N-1}X_ke^{j2\pi kn/N},\quad 0 \leq k \leq N-1,
\end{equation}
where $j = \sqrt{-1}$ and $n$ is the discrete-time index. It is well-known that the OFDM time-domain signal has high peak-to-average power ratio (PAPR) \cite{baxley2009}, which is defined as
\begin{eqnarray}
\label{eq:PAPR}
\text{PAPR} = \frac{\underset{0 \leq n \leq N-1}{\max} |x[n]|^2}{E[|x[n]|^2]},
\end{eqnarray}
where $E[\cdot]$ stands for the statistical expectation. For a large $N$, $\{x[n]\}_{n=0}^{N-1}$ are asymptotically independent and approximately complex Gaussian distributed, and the real part and the imaginary part of $x[n]$ are asymptotically independent \cite{Ochiai2001}. Then, the complementary cumulative distribution
function (CCDF) of the PAPR can be shown to be \cite{Bauml1996a, Ochiai2001}
\begin{equation}
Pr\{\text{PAPR} > r\} = 1- (1-e^{-r})^N.
\end{equation}

%Since Nyquist rate samples might not represent the peaks of the continuous-time signal, it is desirable to show PAPR performance on over-sampled discrete-time signals [4], [5], [16], [18], [20]. It is typical to use an over-sampling factor of $L = 4$ so that the PAPR before the digital to analog (D/A) conversion can accurately describe the PAPR after the D/A conversion.
%\begin{figure}[!t]
%\label{fig_pa}
%  \centering
%  \includegraphics[width=5.5cm]{pafig}
%\caption{Input-output relationship of an ideal linear PA.}
%\end{figure}
%PAs are peak power limited devices. Fig.$\,$1 shows the input-output characteristic of an ideal linear PA \cite{Baxley2008}.

\section{PAPR in VLC-OFDM}
\label{sec:format}

In VLC systems, intensity modulation (IM) is employed at the transmitter. The forward signal drives the LED which in turn converts the magnitude of the input electric signal into optical intensity. The human eye cannot perceive fast-changing variations of the light intensity, and only responds to the average light intensity.  Direct detection (DD) is employed at the receiver. A photodiode (PD) transforms the received optical power into the amplitude of an electrical signal. 

%\begin{figure}[!t]
%  \centering
%  \includegraphics[width=6.5cm]{ledfig2}
%\caption{Ideal linear LED characteristic.}
%\label{fig_led}
%\end{figure}

%\begin{figure*}[!t]
%\centering
%\includegraphics[width=5.2in]{ofdmmodel2}
%\caption{OFDM system model in visible light communications}
%\label{fig_ofdmmodel}
%\end{figure*}
IM/DD schemes require the baseband signal in the VLC to be real-valued. Thus, complex-valued RF-OFDM in (\ref{eq:OFDM}) cannot be used in VLC directly.  
%Fig. \ref{fig_ofdmmodel} shows the OFDM system model in visible light communications. 
According to the property of the inverse Fourier transform, a real-valued
time-domain signal $x[n]$ corresponds to a frequency-domain signal
$X_k$ that is Hermitian symmetric; i.e., $X_k = X_{N-k}^*, 1\leq k \leq N-1$,  
%\begin{eqnarray}
%\label{eq_herm}
%X_k &=& X_{N-k}^*, \quad 1\leq k \leq N-1,
%\end{eqnarray}
where $*$ denotes complex conjugate. The $0$th and $N/2$th subcarrier are null; i.e., $X_0 = 0$, $X_{N/2} = 0$. Then we can obtain the real-valued time-domain signal
$x[n]$ as
\begin{eqnarray}
x[n] &{}={}&\frac{2}{\sqrt{N}}\sum_{k=1}^{N/2-1}\bigg(\Re(X_k)\cos\left(\frac{2\pi
kn}{N}\right)-\\\notag
&&\Im(X_k)\sin\left(\frac{2\pi
kn}{N}\right)\bigg), n = 0, 1,\dots, N-1,
\end{eqnarray}
where $\Re(\cdot)$ denotes the real part of $X_k$, and $\Im(\cdot)$ denotes the imaginary part of $X_k$. Since the DC component is zero ($X_0 = 0$), $x[n]$ has zero mean. According to the Central Limit Theorem, for a large $N$, $\{x[n]\}_{n=0}^{N-1}$ are asymptotically independent and approximately Gaussian distributed \cite{Yu2003} with
probability density function (PDF) $Pr\left(x[n]=z\right)=
1/\sigma_x\cdot\phi\left(z/\sigma_x\right)$,
%\begin{equation}
%Pr\left(x[n]=z\right)=
%\frac{1}{\sigma_x}\phi\left(\frac{z}{\sigma_x}\right),
%\end{equation}
where $\phi(x)= \frac{1}{\sqrt{2\pi}}e^{-\frac{1}{2}x^2}$ is the PDF of the standard Gaussian distribution, and $\sigma_x^2$ is the variance of $x[n]$. According to the definition of PAPR in (\ref{eq:PAPR}), the PAPR of real-valued OFDM signal $x[n]$ is given by
\begin{equation}
\label{eq:realPAPR}
\mathcal{P} =  \frac{\underset{0 \leq n \leq N-1}{\max} x^2[n]}{\sigma_x^2}.
\end{equation}
Assume $x[n]$ is independent and identically distributed, the CCDF of $\mathcal{P}$ can be shown as \cite{Yu2003}
\begin{eqnarray}
&{}{}&\text{CCDF}_{\mathcal{P}}\{r_p\} = Pr\{\mathcal{P} > r_p\}\\\notag
&{}={}& 1 - Pr\{\mathcal{P} \leq r_p\}\\\notag
&{}={}& 1- Pr\{-\sigma_x\sqrt{r_p}\leq x[n] \leq \sigma_x\sqrt{r_p}, 0\leq n \leq N-1\}\\\notag
&{}={}& 1- [Pr\{-\sigma_x\sqrt{r_p}\leq x[0] \leq \sigma_x\sqrt{r_p}\}]^N\\\notag
&{}={}& 1- [\Phi(\sqrt{r_p})-\Phi(-\sqrt{r_p})]^N,
\end{eqnarray}
where $\Phi(x)=\int_{-\infty}^x\phi(t)dt$.

\begin{figure}[!t]
\centering
\includegraphics[width=8cm]{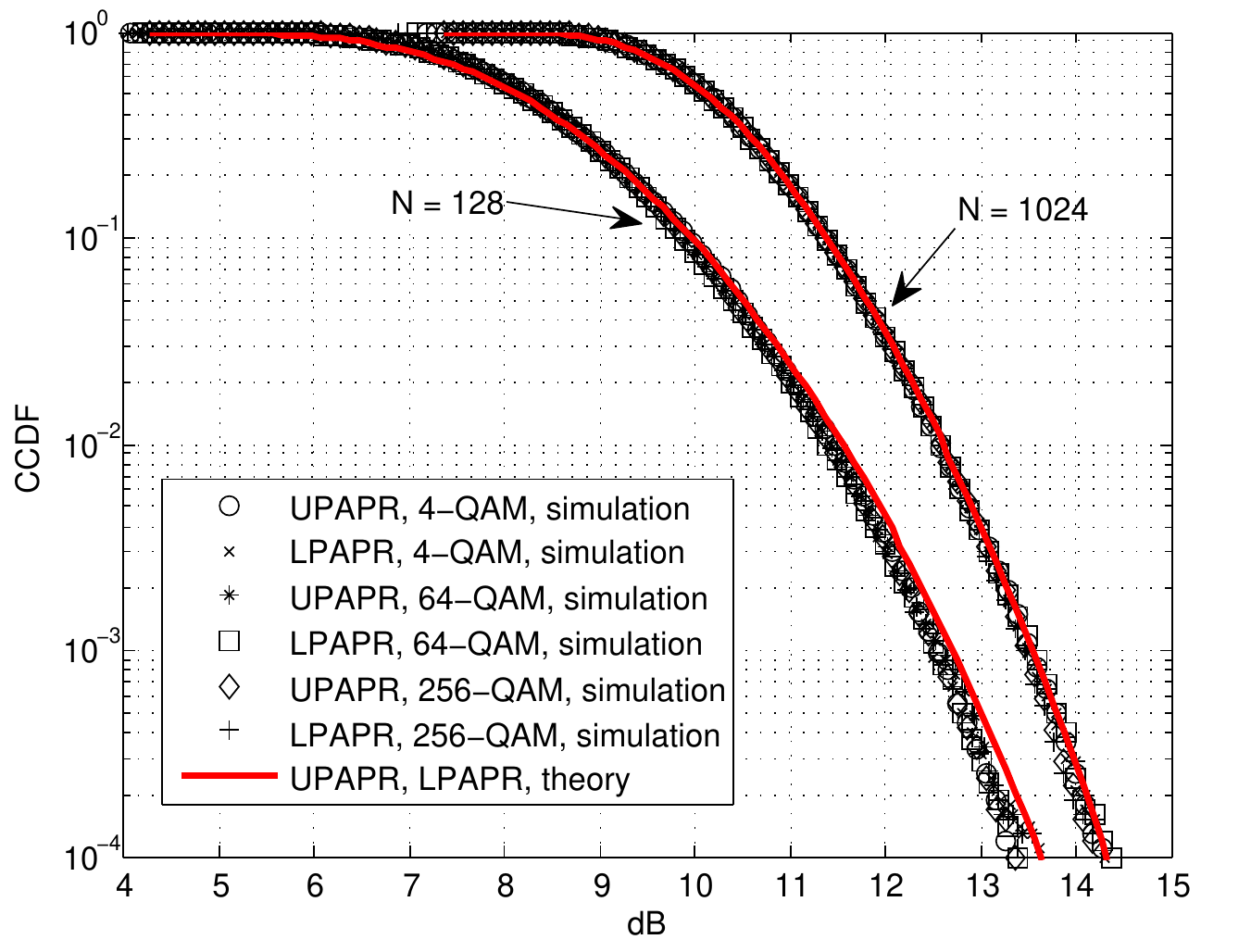}
\caption{CCDF of UPAPR and LPAPR with various constellation orders and numbers of subcarriers.}
\label{fig_papr}
\end{figure}

For the real-valued bipolar signal $\{x[n]\}_{n=0}^{N-1}$, the square of the maximum value  $\left(\underset{0 \leq n \leq N-1}{\max} x[n]\right)^2$ can be seen as the upper peak power, and the square of the minimum value $\left(\underset{0 \leq n \leq N-1}{\min} x[n]\right)^2$ can be seen as the lower peak power. Let us define the upper PAPR (UPAPR) of $x[n]$ as 
%\begin{equation}
%\mathcal{U} \triangleq  \frac{\left(\underset{0 \leq n \leq N-1}{\max} x[n]\right)^2 }{\sigma_x^2},
%\end{equation}
\begin{equation}
\mathcal{U} \triangleq  \left(\underset{0 \leq n \leq N-1}{\max} x[n]\right)^2 /\sigma_x^2,
\end{equation}
%$\mathcal{U} \triangleq  \left(\underset{t \in (0,T]}{\max} x[n]\right)^2 / \sigma_x^2$, 
and define the lower PAPR (LPAPR) of $x[n]$ as %$\mathcal{L} \triangleq  \left(\underset{t \in (0,T]}{\min} x[n]\right)^2 / \sigma_x^2$.
%\begin{equation}
%\mathcal{L} \triangleq  \frac{\left(\underset{0 \leq n \leq N-1}{\min} x[n]\right)^2 }{\sigma_x^2}.
%\end{equation}
\begin{equation}
\mathcal{L} \triangleq  \left(\underset{0 \leq n \leq N-1}{\min} x[n]\right)^2/\sigma_x^2.
\end{equation}
Accordingly, we can derive the CCDF of UPAPR as
\begin{eqnarray}
&{}{}&\text{CCDF}_{\mathcal{U}}\{r_u\} = Pr\{\mathcal{U} > r_u\}\\\notag
&{}={}& 1 - Pr\{\mathcal{U} \leq r_u\}\\\notag
&{}={}& 1- Pr\{x[n] \leq \sigma_x\sqrt{r_u}, 0\leq n \leq N-1\}\\\notag
&{}={}& 1- [Pr\{x[0] \leq \sigma_x\sqrt{r_u}\}]^N\\\notag
&{}={}& 1- \Phi^N(\sqrt{r_u}),
\end{eqnarray}
%\begin{eqnarray}
%\text{CCDF}_{\mathcal{U}}\{r_u\} = Pr\{\mathcal{U} > r_u\} =  1- \Phi^N(\sqrt{r_u}),
%\end{eqnarray}
and the CCDF of LPAPR is derived as
\begin{eqnarray}
&{}{}&\text{CCDF}_{\mathcal{L}}\{r_l\} = Pr\{\mathcal{L} > r_l\}\\\notag
&{}={}& 1 - Pr\{\mathcal{L} \leq r_l\}\\\notag
&{}={}& 1- Pr\{x[n] \geq -\sigma_x\sqrt{r_l}, 0\leq n \leq N-1\}\\\notag
&{}={}& 1- [Pr\{x[0] \geq -\sigma_x\sqrt{r_l}\}]^N\\\notag
&{}={}& 1- \Phi^N(\sqrt{r_l}).
\end{eqnarray}
%\begin{eqnarray}
%\text{CCDF}_{\mathcal{L}}\{r_l\} = Pr\{\mathcal{L} > r_l\} =  1- \Phi^N(\sqrt{r_l}),
%\end{eqnarray}
Note that the UPAPR and the LPAPR have the same CCDF.  Fig. \ref{fig_papr} shows the simulated CCDF of UPAPR and LPAPR with various constellation orders and numbers of subcarriers. In this paper, all the simulation results are taken from 100000 OFDM symbols with 4-QAM, 64-QAM, and 256-QAM constellations. Theoretical results are plotted as well to examine our analysis. We can observe that the distribution of $\mathcal{U}$ and $\mathcal{L}$ are independent of the constellations orders and increase with more subcarriers. When $N = 128$, slight differences can be found between the simulated results and the theoretical values. When $N = 1024$, the simulated results match the theoretical values very well because the central limit theory holds better. 

From the definitions of UPAPR and LPAPR, we can see that UPAPR and LPAPR are not independent distributed. Therefore, it is necessary to know the joint cumulative distribution function of UPAPR and LPAPR, which can be obtained as
\begin{eqnarray}
&{}{}&F_{\mathcal{L}, \mathcal{U}}(r_l,r_u) = Pr\{\mathcal{L} \leq r_l, \mathcal{U} \leq r_u\} \\\notag
&{=}& Pr\{-\sigma_x\sqrt{r_l} \leq x[n] \leq \sigma_x\sqrt{r_u}, 0\leq n \leq N-1\}\\\notag
&{=}& [Pr\{-\sigma_x\sqrt{r_l} \leq x[0] \leq \sigma_x\sqrt{r_u}]^N\\\notag
&{=}& [\Phi(\sqrt{r_u}) - \Phi(-\sqrt{r_l})]^N.
\end{eqnarray}
%\begin{eqnarray*}
%&{}{}&F_{\mathcal{L}, \mathcal{U}}(r_l,r_u) = Pr\{\mathcal{L} \leq r_l, \mathcal{U} \leq r_u\} \\\notag
%&{=}& [\Phi(\sqrt{r_u}) - \Phi(-\sqrt{r_l})]^N
%\end{eqnarray*}
The joint PDF of UPAPR and LPAPR is
\begin{eqnarray}
\label{eq:jointpdf}
&{}{}&f_{\mathcal{L}, \mathcal{U}}(r_l,r_u)\\\notag
&{=}& \frac{\partial^2F_{\mathcal{L}, \mathcal{U}}(r_l,r_u)}{\partial r_l\partial r_u}\\\notag
&{=}& \frac{\phi(\sqrt{r_u})\phi(\sqrt{r_l})}{4\sqrt{r_lr_u}}N(N-1)[\Phi(\sqrt{r_u}) - \Phi(-\sqrt{r_l})]^{N-2}.
\end{eqnarray}
%\begin{eqnarray*}
%&&f_{\mathcal{L}, \mathcal{U}}(r_l,r_u) = Pr\{\mathcal{L} = r_l, \mathcal{U}= r_u\}\\\notag
%&=&\frac{\phi(\sqrt{r_u})\phi(\sqrt{r_l})}{4\sqrt{r_lr_u}}N(N-1)[\Phi(\sqrt{r_u}) - \Phi(-\sqrt{r_l})]^{N-2}
%\end{eqnarray*}
\section{Linear scaling and biasing}
LEDs place dynamic range constraints $[I_L, I_H]$ on the input signal, where $I_L$ denotes turn-on current and $I_H$ denotes the maximum input current \cite{Elgala2009}. The Dynamic range can be denoted by $D \triangleq  I_H-I_L$. Since $I_L$ is positive, the bipolar signal $x[n]$ cannot serve as the input of LEDs directly. The forward signal $y[n]$ can be obtained from the OFDM signal $x[n]$ after both a linear scaling and a biasing operation; i.e., 
\begin{equation}
\label{eq:linmodel}
y[n] = \alpha x[n] + B,  \quad 0 \leq n \leq N-1, 
\end{equation}
where $\alpha$ denotes the scaling factor used to control the input power back-off 
\begin{equation}
\label{eq:boff}
\gamma \triangleq \frac{D^2}{\alpha^2\sigma_x^2},
\end{equation}
%$\gamma \triangleq D^2/(\alpha\sigma_x)^2$, 
and $B$ denotes the biasing level. $\alpha$ and $B$ are both real-valued. The resulting signal, $y[n]$, has a mean value $B$ and a variance $\sigma_y^2 = E[\alpha^2]\sigma_x^2$. 

Brightness control is essential for the illumination function in VLC. The principle of brightness control is to make the average amplitude of the input forward signal equal to $I_{\mathrm{des}}$, which corresponds to a desired average output optical intensity. 
%To avoid flicker of LED, it is better to keep the average value of each N-length symbol $y[n]$ constant.  
Since the mean value of the input forward signal signal $y[n]$ is equal to $B$, it is straightforward to set biasing level $B$ equal to $I_{\mathrm{des}}$; i.e., $B = I_{\mathrm{des}}$, which is called the biasing adjustment method \cite{Yu2013a}. Therefore, the biasing level $B$ is determined by the illumination requirements. Let us define biasing ratio as 
%$\varsigma \triangleq (B-I_{L})/D$.
\begin{equation}
\label{eq:bratio}
\varsigma \triangleq \frac{B-I_{L}}{D}.
\end{equation}
Without loss of generality, we only consider biasing ratio in the range $0 \leq \varsigma \leq 0.5$, because any forward signal $s[n]$ with  biasing ratio $0.5 < \varsigma \leq 1$ can be created from $y[n]$, which has biasing ratio $0 \leq 1-\varsigma \leq 0.5$ and is within the dynamic range $[I_L, I_H]$, by $s[n] = I_H+I_L - y[n]$.
\subsection{Symbol-invariant scaling factor}
\begin{figure}[!t]
  \centering
  \includegraphics[width=7.5cm]{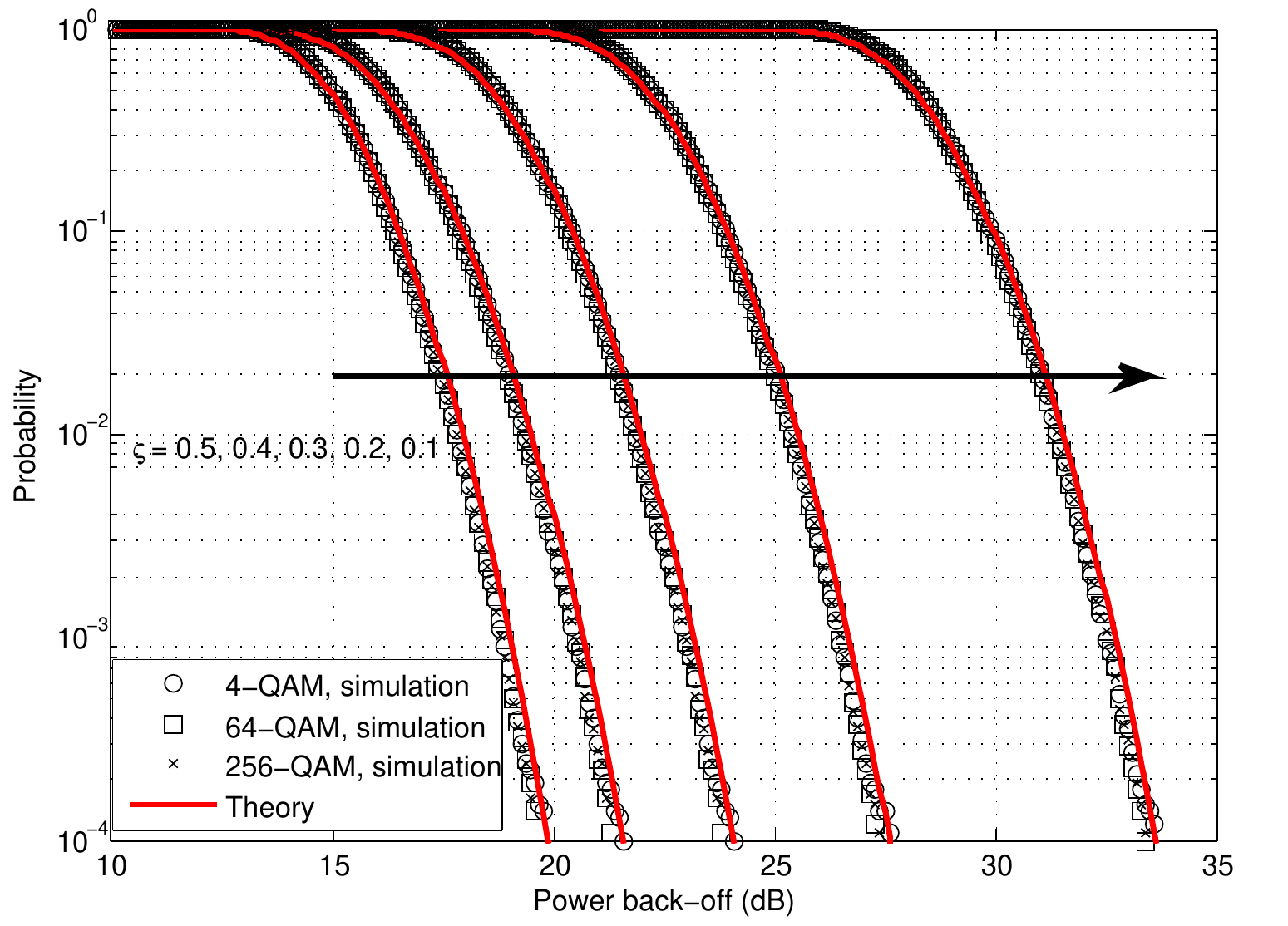}
\caption{Probability that the input symbol $\{y[n]\}_{n=0}^{N-1}$ is beyond the dynamic range of LEDs given power back-off and biasing ratio (128 subcarriers).}
\label{fig_prob2}
\end{figure}

\begin{figure}[!t]
  \centering
  \includegraphics[width=7.4cm]{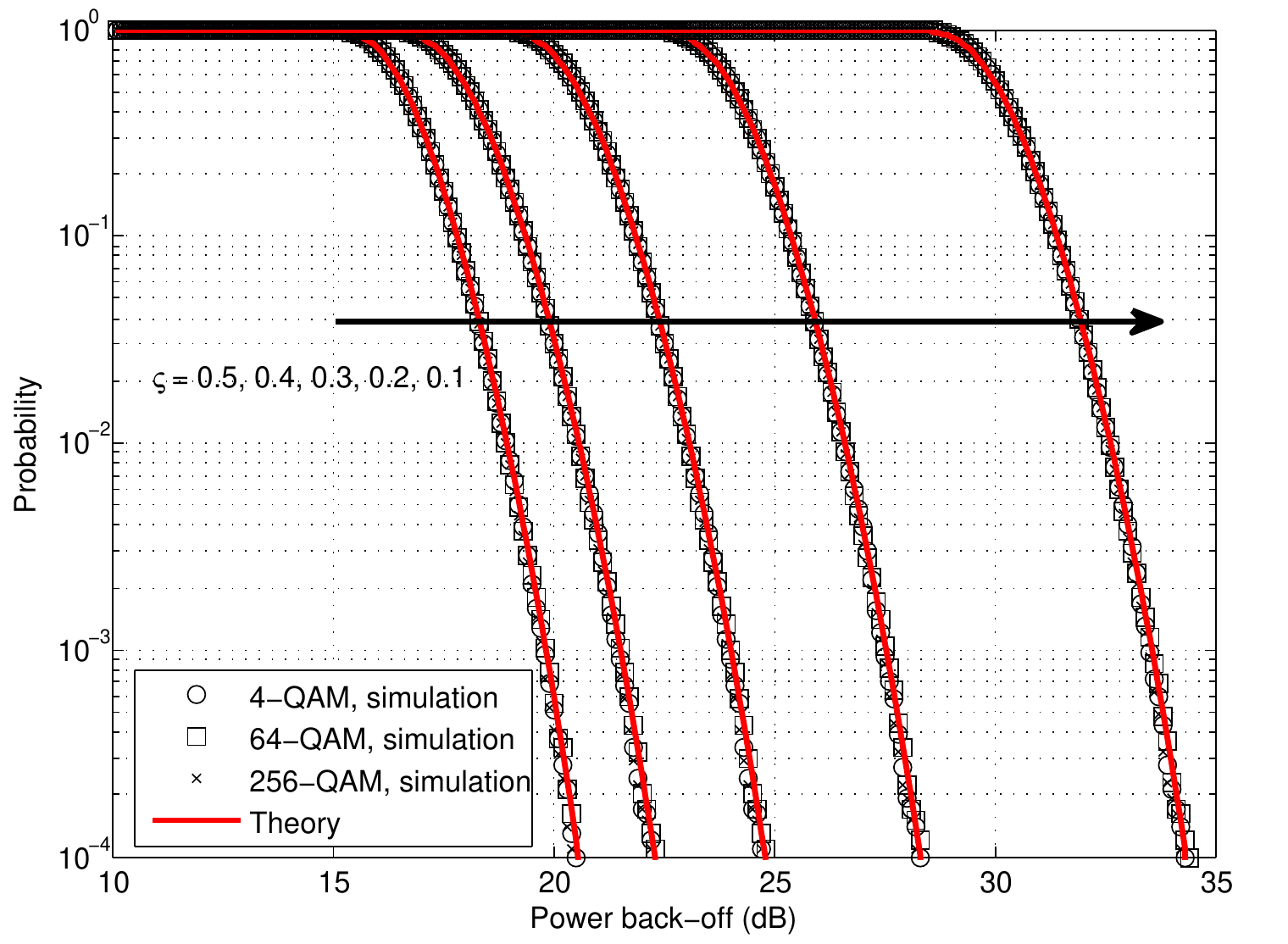}
\caption{Probability that the input symbol $\{y[n]\}_{n=0}^{N-1}$ is beyond the dynamic range of LEDs given power back-off and biasing ratio (1024 subcarriers).}
\label{fig_prob1}
\end{figure}

Assume that the scaling factor $\alpha$ or input power back-off $\gamma$ is fixed for all OFDM symbols. Given the biasing ratio $\varsigma$ and the input power back-off $\gamma$, it is useful to know the probability that the input OFDM symbol $\{y[n]\}_{n=0}^{N-1}$ is beyond the dynamic range of LEDs, which is given by
\begin{eqnarray}
\label{eq:probviol}
&& Pr\{\{y[n]\}_{n=0}^{N-1} \text{ is out of dynamic range}\mid \gamma, \varsigma\}\\\notag
&=& Pr\{y[n]> I_H, \text{or}\,\, y[n] < I_L\mid \gamma, \varsigma\}\\\notag
&=& Pr\left\{x[n] > \frac{I_H - B}{\alpha}, \text{or}\,\, x[n] < \frac{I_L - B}{\alpha}\mid \gamma, \varsigma\right\}\\\notag
&=& 1 - Pr\left\{\frac{I_L - B}{\alpha} \leq \{x[n]\}_{n=0}^{N-1} \leq \frac{I_H - B}{\alpha} \mid \gamma, \varsigma\right\}\\\notag
&=& 1 - Pr\left\{\frac{I_L - B}{\alpha\sigma_x} \leq \frac{\{x[n]\}_{n=0}^{N-1}}{\sigma_x} \leq \frac{I_H - B}{\alpha\sigma_x} \mid \gamma, \varsigma\right\}\\\notag
&=& 1 - Pr\left\{-\varsigma\sqrt{\gamma} \leq \frac{\{x[n]\}_{n=0}^{N-1}}{\sigma_x} \leq (1-\varsigma)\sqrt{\gamma}, \right\}\\\notag
&=& 1 - Pr\{\mathcal{L} \leq \varsigma^2\gamma, \text{and}\,\,\mathcal{U} \leq (1-\varsigma)^2\gamma\}\\\notag
&=& 1 - F_{\mathcal{L}, \mathcal{U}}(\varsigma^2\gamma,(1-\varsigma)^2\gamma),
\end{eqnarray} 
where $(I_L - B)/\alpha\sigma_x = -\varsigma\sqrt{\gamma}$ and $(I_H - B)/\alpha\sigma_x=(1-\varsigma)\sqrt{\gamma}$ are obtained from the Eqs. (\ref{eq:boff}) (\ref{eq:bratio}). We can see that the probability in Eq. (\ref{eq:probviol}) depends on the joint distribution of UPAPR and LPAPR. Fig. \ref{fig_prob2} shows simulated and theoretical results for 128 subcarriers. Fig. \ref{fig_prob1} shows simulated and theoretical results for 1024 subcarriers. We can observe that the joint distribution of UPAPR and LPAPR is independent of constellations, and the simulated results match the theoretical values well.  For a biasing ratio $ \varsigma \in [0,0.5]$, lower biasing ratio requires larger input power back-off to achieve the same probability.  

\subsection{Symbol-variant scaling factor}

\begin{figure}[!t]
  \centering
  \includegraphics[width=7.6cm]{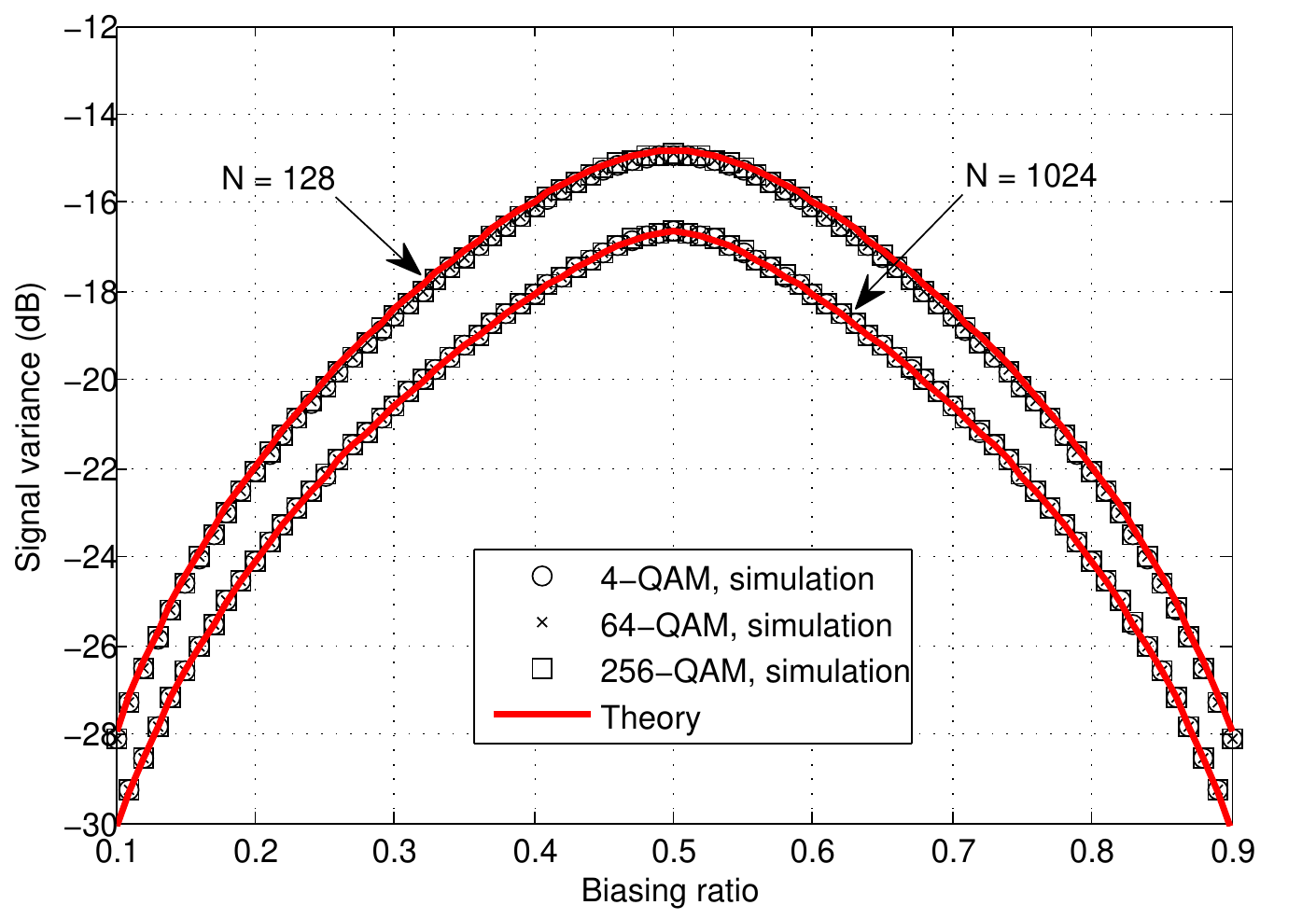}
\caption{Variance $\sigma_y^2$ as a function of the biasing ratio with normalized dynamic range.}
\label{fig_var}
\end{figure}

Assume that the scaling factor $\alpha$ can be adjusted symbol by symbol. The variance $\sigma_y^2$ can be maximized by selecting $\alpha$ with the greatest value for each OFDM symbol. To ensure $y[n]$ is within the dynamic range of the LED, we can obtain an $\alpha$ with the greatest value as
\begin{equation}
\label{eq_alphap}
\alpha = \min \left\{ \frac{I_H - B}{\underset{0 \leq n \leq N-1}{\max} x[n]}, \frac{I_L - B}{\underset{0 \leq n \leq N-1}{\min} x[n]} \right\}
\end{equation}
We can obtain the variance of $y[n]$ as
\begin{eqnarray}
\label{eq:var}
&&\sigma_y^2 = \sigma_x^2E[\alpha^2]\\\nonumber
 &=& D^2 E_{\mathcal{U},\mathcal{L}}\left[\min \left\{\frac{(1-\zeta)^2}{\mathcal{U}}, \frac{\zeta^2}{\mathcal{L}} \right\}\right]\\\notag
 &=&D^2\int_{0}^{\infty}\int_{0}^{\infty}\min \left\{\frac{(1-\zeta)^2}{r_u}, \frac{\zeta^2}{r_l} \right\}f_{\mathcal{L}, \mathcal{U}}(r_l,r_u)dr_ldr_u,
 %&\approx&D^2\sum\sum\min\left\{\frac{(1-\zeta)^2}{r_u}, \frac{\zeta^2}{r_l} \right\}f_{\mathcal{L}, \mathcal{U}}(r_l,r_u),
\end{eqnarray}
where $f_{\mathcal{L}, \mathcal{U}}(r_l,r_u)$ is the joint PDF of UPAPR and LPAPR from Eq. (\ref{eq:jointpdf}).  We can observe that the variance $\sigma_y^2$ depends on three factors: biasing ratio, upper PAPR of the OFDM signal and lower PAPR of the OFDM signal. Note that the dynamic range $D$ is a fixed value, which is determined by characteristics of LEDs. The scaling factor $\alpha$ varies symbol by symbol since $\mathcal{U}$ and $\mathcal{L}$ are both random variables.  We treat $\alpha$ as part of the channel and assume that $\alpha$ for each symbol can be perfectly estimated at the receiver. Fig. \ref{fig_var} is a plot of the variance $\sigma_y^2$ versus the biasing ratio with normalized dynamic range. The plots demonstrate that our theoretical analysis matches the simulated results very well. The variances are identical for all three constellation orders since the distributions of LPAPR and UPAPR are independent of constellations. The variance decrease with increasing subcarriers because both the UPAPR and LPAPR will increase when there are more subcarriers. Since the OFDM signal has a symmetric distribution, the maximum variance occurs at biasing ratio 0.5 when the OFDM signal is biased around the middle point of the dynamic range.

\section{Conclusions}
In this paper, we derived the CCDF and the joint distribution of UPAPR and LPAPR. The performance of VLC-OFDM with dynamic range and average amplitude constraints are shown to be directly related with UPAPR and LPAPR. Simulation results matched the theoretical analysis well.

%\vfill
%\pagebreak

\vfill\pagebreak

\bibliographystyle{IEEEbib}
\bibliography{E:/bib/library}

\end{document}